\documentclass[%
 twocolumn,
 amsmath,amssymb,
 aps,
]{revtex4-2}

\usepackage{graphicx}
\usepackage{changepage}
\usepackage{dcolumn}
\usepackage{xcolor}
\usepackage{bm}
\usepackage{enumitem}
\usepackage{placeins}
\usepackage{tikz}
\usetikzlibrary{positioning, shapes, backgrounds, fit}
\usetikzlibrary{positioning, arrows.meta, decorations.pathreplacing}
\usepackage{hyperref}

\usepackage[normalem]{ulem}

\begin{document}

\preprint{APS/123-QED}

\title{Virality detection and control strategies in rumor models}%
\author{Eva Rifà}

\email{eva.rifa@eurecat.org}
\affiliation{%
 Eurecat, Centre Tecnològic de Catalunya, Barcelona, Spain
}%

\affiliation{%
 Facultat de Física, Universitat de Barcelona (UB), Barcelona, Spain
}%

\author{Julian Vicens}

\affiliation{
 Eurecat, Centre Tecnològic de Catalunya, Barcelona, Spain
}%
\author{Emanuele Cozzo}
\affiliation{Universitat de Barcelona Institute of Complex Systems (UBICS), Universitat de Barcelona, Barcelona, Spain}
\affiliation{Departament de Física de la Matèria Condensada, Universitat de Barcelona, Barcelona, Spain}
\affiliation{CNSC-IN3, Universitat Oberta de Catalunya, Barcelona, Spain}

\date{\today}

\begin{abstract}
We study the dynamics and intervention strategies of a rumor using the modified Maki-Thompson model. A key challenge in social networks is distinguishing between natural increases in transmissibility and artificial injections of rumor spreaders, such as through broadcast events or astroturfing. Using stochastic simulations, we compare two scenarios: one with organic growth in transmissibility and another with externally injected spreaders. Although both lead to high autocorrelation, only the organic growth produces oscillatory patterns in autocorrelation at multiple lags, an effect we can analytically explain using the $N$-intertwined mean-field approximation. This distinction offers a practical tool to identify the origin of rumor virality and also infer its transmissibility. Our approach is validated analytically and tested on real-world data from Twitter during the announcement of the Higgs boson discovery. In addition to detection, we also explore control strategies. We show that the average lifetime of a rumor can be manipulated through targeted interventions: placing spreaders at specific locations in the network. Depending on their placement, these interventions can either extend or shorten the lifespan of the rumor.
\end{abstract}

\maketitle

\section{Introduction}
\label{sec1}

In the digital era, social networks and platforms have become central elements in the dissemination of information, influencing the dynamics of public discourse and communication. It has long been recognized that as long as dominant social media platforms are owned by monopolistic companies whose primary goal is profit maximization, they cannot serve as spaces where rumor propagation is truly spontaneous. That is why virality, understood as the widespread reproduction of a message, is not a neutral or uniform process.
Furthermore, these platforms have enabled mechanisms of persuasion in the battle over public opinion, including disinformation campaigns and the emergence of covert manipulation tactics such as astroturfing \cite{viralitatdelmal_2024}.

Astroturfing refers to the practice of simulating the appearance of a grassroots movement through the mass mobilization of fake or automated accounts \citep{viralitatdelmal_2024}. While this practice predates the internet, its current scale, amplified by platforms such as X (formerly Twitter), makes it cheaper, more global, and more difficult to detect \citep{arce-garcia_tipos_2023}. It has been used in a variety of contexts, including  aggressive marketing strategies and electoral manipulation. The Internet Research Agency's campaigns during the 2016 U.S. presidential elections \citep{bail_assessing_2020} are a well-known example, but recent studies have documented similar operations in South Korea \citep{keller_political_2020} and Thailand \citep{sirikupt_drowning_2024}, among others \citep{wang_evidence_2023}. Since ground truth is rarely available \citep{keller_political_2020}, researchers suggest focusing on identifying coordinated behavior rather than specific content \citep{schoch_coordination_2022}. Astroturfing must be driven by the interests of economic or political power. When this is not the case, we can more generally speak of an external impulse to the system. For example, the diffusion through traditional media (television, radio, streaming) can trigger similar dynamics without manipulative intent \citep{gelper_effect_2024, brannon_speed_2024}. These dynamics allow us to delve deeper into the study of viral phenomena and develop strategies to manage the propagation of information online.

One of the most extensively studied models for rumor spreading is the Maki-Thompson model \citep{maki_mathematical_1973}, which, similar to epidemiological models, describes individuals transitioning between states of ignorance, spreading, and stifling depending on their interactions. For many years it was believed that this model would always have a nonzero probability of reaching a non-vanishing fraction of the population \citep{barrat_dynamical_2008}, and therefore that it did not exhibit a phase transition. However, a recent study by \citet{ferraz_de_arruda_subcritical_2022} demonstrated the existence of a critical phase transition. This finding, which arises from accounting for node-to-node correlations neglected in mean-field approaches, challenged long-standing assumptions and opened new directions for research in rumor dynamics.

The existence of a critical transition in the rumor model has important implications. In particular, it enables the study of early warning signals (EWS), statistical indicators that detect when systems are approaching a tipping point and experiencing critical slowing down (CSD). EWS have been successfully applied in fields such as epidemiology \citep{oregan_theory_2013}, ecology and climate science \citep{george_early_2023}, while, as far as we know, they have not yet been explored in the context of rumor dynamics. Moreover, while in recent years intervention strategies and forecasting have become standard tools in epidemic modeling \citep{ullon_controlling_2023}, only a few studies have taken a similar approach to investigate the spread of rumors \citep{belen_impulsive_2005, pearce_impulsive_2005}, that do not account for the system's critical dynamics.

The current context, together with theoretical advances in rumor spreading modeling, leads us to ask the following questions: Can we distinguish whether the rumor virality is driven by an external impulse (such as a broadcast or astroturfing campaign) or by an internal increase in transmissibility? And what are the optimal strategies to control the reach of a rumor? Answering these questions is not only theoretically relevant but also brings significant practical implications. Distinguishing between the organic growth of rumors and coordinated external behavior is crucial for identifying disinformation campaigns and preserving the integrity of online discourse. Furthermore, by identifying the optimal way for controlling the rumor spread, can serve as a basis for designing robust communication strategies to amplify beneficial information and mitigate harmful content on digital platforms. 
This article is organized as follows: Section \ref{model_definition} presents the model, Section \ref{results} shows the results and Section \ref{discussion} provides the discussion.

\section{Model definition}
\label{model_definition}

We consider the modified version of the Maki-Thompson (MT) rumor spreading model \cite{ferraz_de_arruda_subcritical_2022}, where each individual (node \(i\)) can be in one of three states: ignorant ($X_i=1$), spreader ($Y_i=1$), or stifler ($Z_i=1$) where $X_i+Y_i+Z_i=1$. These are modeled as Bernoulli random variables taking values in $\{0,1\}$. Transitions between states occur as independent Poisson processes on an undirected network defined by adjacency matrix $A=(A_{ij})$. The transition rules are as follows:
\begin{align*}
Y + X &\xrightarrow{\lambda} Y + Y  &\text{(spreading)}\\
Y + Y &\xrightarrow{\alpha} Z + Y &\text{(stifling by another spreader)}\\
Y + Z &\xrightarrow{\alpha} Z + Z &\text{(stifling by a stifler)}\\
Z &\xrightarrow{\delta} X &\text{(forgetting)}
\end{align*}
The difference with an epidemiological SIRS (Susceptible-Infected-Recovered-Susceptible) model resides in the removal of spreaders. While in the last the transition is spontaneous, in the former it depends on the contacts. 
With respect to the original MT model \citep{maki_mathematical_1973}, in which $\delta=0$ and stiflers never become susceptible to be spreaders again, in the modified MT model, $\delta>0$ introduces a spontaneous transition from stifler to ignorant. Thus, while the original model has an infinite number of absorbing states, namely any state with $Y_i=0 \ \forall i$, the modified model has a unique (rumor-free) absorbing state ($X_i=1\ \forall i$).

As showed in \cite{ferraz_de_arruda_subcritical_2022}, a phase transition exists both in the MT original and modified model. In the original MT model, the critical point separates a regime in which the number of stiflers when the process reaches one of the infinite absorbing states does not scale with the system size, from one in which it diverges with the system size. In the modified model, the critical point separates a subcritical regimes in which the process ends in the absorbing state, from a regime in which it stays in an active state. The lifetime of a rumor diverges at the critical point. However, while in epidemiological models the lifetime decays exponentially in the subcritical regime, in considered rumor models has a non monotonic power-law behavior (see Fig. \ref{fig_autocor_lifetime_qs}).

The system evolves as a continuous-time Markov chain on the constrained state space  \(\{0,1\}^{3N}\), with transition rates determined by pairwise interactions. Thus, it admit in principle an exact analysis. However, it becomes intractable for large $N$, so we consider a first-order approximation by taking expectations of the indicator variables:
\begin{equation}
\label{MF_eqs}
    \left\{\begin{array}{ll}\frac{dx_i}{dt} = \delta z_i - \lambda \sum^N_{k = 1}A_{ki}x_iy_k\\
     \frac{dy_i}{dt} = \lambda \sum^N_{k = 1}A_{ki}x_iy_k - \alpha \sum^N_{k = 1}A_{ki}y_i(y_k + z_k)\\ 
    \frac{dz_i}{dt} = -\delta z_i + \alpha \sum^N_{k = 1}A_{ki}y_i(y_k + z_k)\end{array}\right.
\end{equation}
where $x_i(t) = \mathbb{E}[X_i(t)]$, $y_i(t)=\mathbb{E}[Y_i(t)]$ and $z_i(t)=\mathbb{E}[Z_i(t)]$. This approximation captures the average behavior of the system while accounting for local network structure via the adjacency matrix.

\section{Results}
\label{results}

First, we study EWS in the rumor model (Section \ref{EWS_section}) and whether they allow us to distinguish organic growth from external injections (Section \ref{organic_injection}). Second, we derive these results analytically (Sections \ref{NIMFA_der} and \ref{analytical_conditions}) and third, we present a case study (Section \ref{case_study}). Finally, we show how targeted interventions can lengthen or shorten the rumor's lifetime (Section \ref{control_strategies}).

\subsection{Analysis of rumor dynamics through early warning signals}
\subsubsection{Autocorrelation patterns of the rumor model}
\label{EWS_section}
We investigate whether EWS can provide insights into the dynamics of rumor spreading. One of the most commonly used statistical indicators for detecting critical transitions is the autocorrelation \citep{oregan_theory_2013, ullon_controlling_2023}, which measures the correlation of a signal with a delayed version of itself. As a system approaches a critical point, its recovery rate from perturbations typically slows down, leading to an increase in autocorrelation. This phenomenon, known as critical slowing down (CSD), makes autocorrelation a powerful tool for anticipating shifts in critical complex systems.

\begin{figure}[h!]
    \centering
    \includegraphics[width=0.8\linewidth]{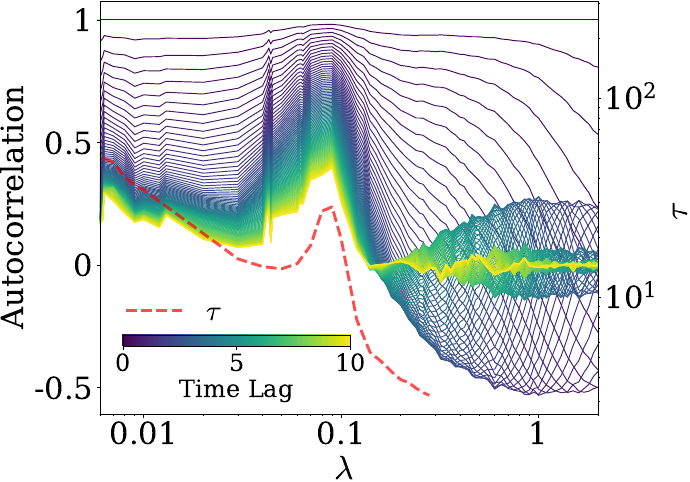}
    \caption{The autocorrelation as a function of the spreading rate ($\lambda$) at multiple lags of the density of spreaders starting from a single spreader. The red dashed line indicates the average lifetime of finite realizations ($\tau$). The autocorrelation is computed from simulations using the quasi-stationary method. The parameters used are $\delta =0.89, \alpha = 0.5$ for a network of size $N = 1000$ and $\langle k\rangle=10$.}
    \label{fig_autocor_lifetime_qs}
\end{figure}

We begin by analyzing the autocorrelation of the density of spreaders at different lags (the time separation between two observations) across varying spreading rates to explore the different regimes of the model. Additionally, we examine the lifetime ($\tau$) \cite{boguna_nature_2013} of the rumor as a marker to identify the critical point ($\lambda_c$) (see Appendix \ref{methods_lifetime} for details). As expected, the autocorrelation increases near the $\lambda_c$, consistent with critical slowing down (see Fig. \ref{fig_autocor_lifetime_qs}).

However, an important subtlety arises in the subcritical regime, where the system evolves very slowly due to the rarity of spreading events. In this regime, autocorrelation remains artificially high, not because the system is approaching a transition, but because it is barely changing. This slow behavior leads to high autocorrelation values that can be misleading if interpreted in isolation. This finding suggests that, in rumor models, autocorrelation should be interpreted in the context of dynamical regime of the system.

\subsubsection{Distinguishing organic growth from an external impulse}
\label{organic_injection}
\begin{figure*}[htb]
  \includegraphics[width=\linewidth]{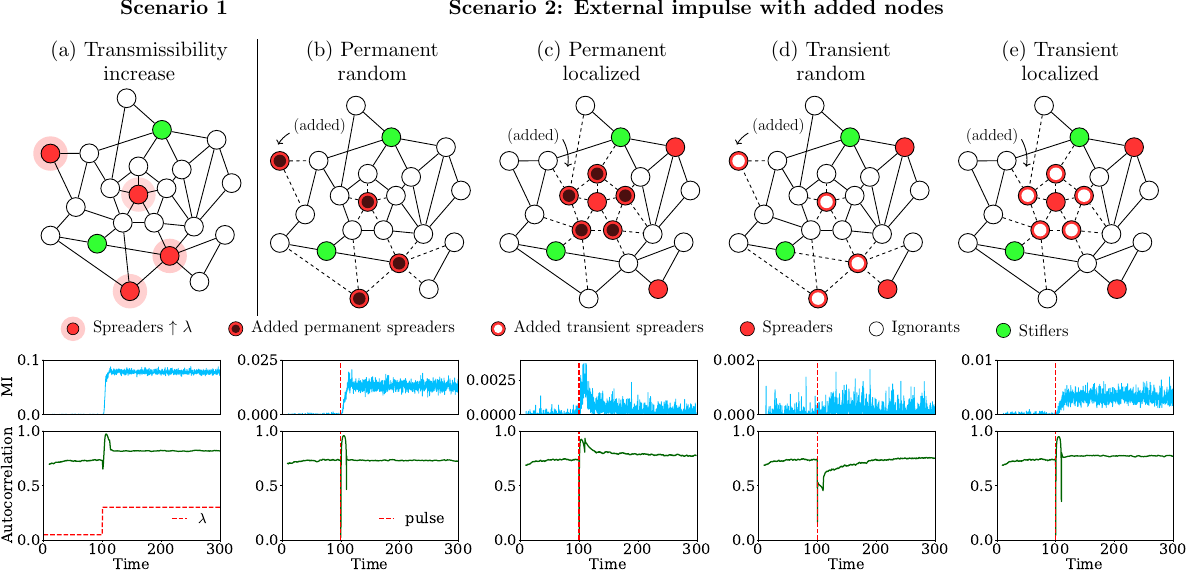}
    \caption{Schematic representation of two scenarios: (1) sudden increase in the transmissibility rate ($\lambda$) and (2) external impulse with fixed spreading rate. Top panels show network representations: solid edges represent existing connections, while dashed edges indicate externally added nodes. In scenario (2), four variants are simulated: (b) a pulse of permanent spreaders added at a random locations; (c) a localized pulse of permanent spreaders added adjacent to an existing spreader; (d) a pulse of transient spreaders (converted to common population) added at  random locations; and (e) a localized pulse of transient spreaders added adjacent to an existing spreader. Bottom panels show the mutual information (MI) and autocorrelation of the spreaders density. Results are averaged over 1000 realizations using a rolling window of 50 time steps. Parameters are: $\alpha = 0.5, \delta = 0.89$, $N = 1000$, $\langle k\rangle = 10$ and the external perturbation is of 50 new spreaders. See Appendix \ref{methods} for simulation details.}
    \label{fig-ews_inst}
\end{figure*}

We now focus on realistic rumor spreading dynamics and investigate whether early warning signals (EWS) can distinguish different types of virality, understood as distinct pathways through which a rumor can reach large fraction of the population. In particular, we examine whether EWS are able to discriminate between externally induced virality and that arising endogenously from the system itself.

To address this question, we designed two scenarios that lead to a similar density of spreaders, but that correspond to different routes towards a viral regime. The first scenario consists of a density pulse that mimics a sudden intrusion of new spreaders interacting with the rumor, a process associated with astroturfing strategies or, more generally, external shocks. The second scenario corresponds to an abrupt increase in the spreading rate, causing the system to transition from a subcritical to a supercritical regime, thus giving rise to endogenous virality emerging from the internal dynamics of the system. In both experiments, we employ two early warning indicators to track changes in the system's dynamics. One of them is the autocorrelation, which we compute from the density of spreaders, capturing the similarity of the system's state across successive time intervals. The other is the mutual information between node states, which quantifies the amount of information obtained about one node by observing the other.

To simulate the first scenario, corresponding to an organic growth of the interest in the rumor, we instantaneously tune the transmissibility parameter from the subcritical to the supercritical regime. The second scenario represents an external impulse and is implemented through an artificial injection of spreaders. In this case, we consider four variants of the injection mechanism, accounting for two independent factors. On the one hand, we distinguish between the activity of the injected spreaders: either permanent spreaders, which remain continuously active, or transient spreaders, which behave like normal agents and follow the same rumor spreading rules. On the other hand, we consider the spatial placement of the injected spreaders, distinguishing between random injection and localized injection near existing spreader nodes. A schematic representation of the scenarios and the corresponding indicators is shown in Fig. \ref{fig-ews_inst}, while further methodological details are provided in Appendix \ref{methods}.

For the transmissibility increase scenario, we observe that the mutual information exhibits a sharp increase at the time of the perturbation and remains high thereafter. Similarly, the autocorrelation displays a peak at the perturbation, followed by a relaxation to a value higher than that observed prior to the transition. In the case of artificially injected spreaders, the perturbation induces an abrupt change in the time series of spreaders density, which is also reflected in the early warning indicators across all injection variants. In particular, the mutual information shows a sudden increase at the perturbation both for the transmissibility increase and for the injection of permanent spreaders. 

The autocorrelation, however, exhibits more heterogeneous behavior depending on the injection mechanism. When permanent spreaders are introduced, a pronounced peak is observed at the perturbation, after which the autocorrelation stabilizes at a higher or lower value depending on whether the injection is random or localized. In contrast, when the injected spreaders behave as normal agents, the autocorrelation takes longer to stabilize, and in the case of localized injection it displays a drop rather than a peak.

The key challenge is therefore to distinguish, based solely on the post-perturbation time series of spreaders, whether the observed change originates from an internal dynamical transition or from an external influence. Our results indicate that this distinction remains ambiguous, as the temporal evolution of the autocorrelation in the two scenarios exhibits qualitatively similar features.

\begin{figure}[h!]
\centering
\includegraphics[width=\linewidth]{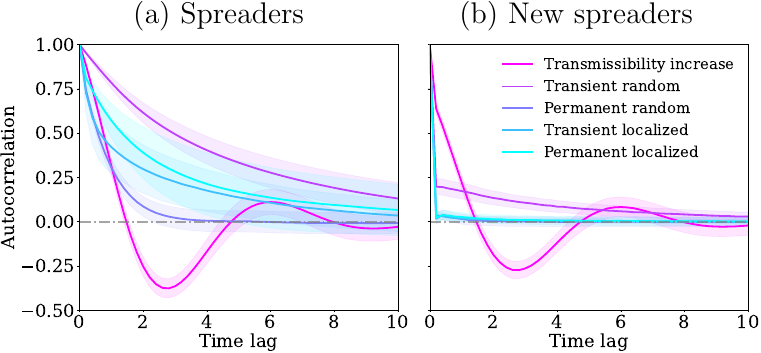}
\caption{
       Autocorrelation of (a) spreaders and (b) new spreaders density as a function of lag after the perturbation for the different scenarios of Fig. \ref{fig-ews_inst}.
        }
\label{fig_lags_new_spreaders}
\end{figure}

To resolve this ambiguity we examine the autocorrelation across lags (Fig. \ref{fig_lags_new_spreaders}a). In the case of a permanent increase in transmissibility that leaves the system in a metastable steady state, correlations decay as damped oscillations; an external injection, by contrast, produces an exponential decay. For the increase in transmissibility, we can see oscillations also when we measure only the time series for the new spreaders density (Fig. \ref{fig_lags_new_spreaders}b), counted as the new infected nodes during each time step. This finding is very useful when working with real data, where it is often difficult to define when a spreader becomes a stifler. Therefore it will be much easier for us to measure the density of new spreaders. Furthermore, oscillations are also visible for different network structures (see Appendix \ref{oscillations_structures}).

\subsubsection{The NIMFA autocorrelation in the steady state}
\label{NIMFA_der}
In this section, we developed an analytical approximation for the autocorrelation in the steady state using the $N$-intertwined mean-field approximation (NIMFA) \cite{4549746, Van_Mieghem_2014, liu_autocorrelation} that allows us to write effective rates ($\tilde{\lambda}_{\infty}$, $\tilde{\alpha}_{\infty}$) for each node transition. 
\begin{align}
    & \tilde{\lambda}_{i\infty} \equiv \lambda \sum^N_{k = 1}A_{ki}y_{i\infty} \\
    & \tilde{\alpha}_{i\infty} \equiv \alpha \sum^N_{k = 1}A_{ki}(y_{k\infty} + z_{k\infty})
\end{align}

The autocorrelation of the spreading state ($V$) of a node $j$ between time $s$ and $t$ is:
\begin{align}
    R_{j}(s,t) \triangleq \frac{E[V_{j}(s)V_{j}(t)]-E[V_{j}(s)]E[V_{j}(t)]}{\sqrt{Var[V_{j}(s)]Var[V_{j}(t)]}}
\end{align}

In a first-order approximation we assume that the probabilities are independent, for Bernoulli random variables $E[V_i(t)V_j(t)] = E[V_i(t)]E[V_j(t)]$. Upon further derivation (see Appendix \ref{analytica_autocor}), the analytical expression for the autocorrelation is obtained:

\resizebox{\columnwidth}{!}{
\begin{minipage}{\columnwidth}
\thinmuskip=0mu
\medmuskip=0mu
\thickmuskip=0mu
\begin{equation}
\begin{split}
&R_{i\infty}(h) = \frac{1}{1-v_{i\infty}} \Biggr[ \frac{1}{\tilde{\alpha}\tilde\lambda+\tilde{\alpha}\delta +\tilde\lambda\delta} \Big[
        \delta\tilde\lambda + \frac{\tilde{\alpha}}{2}(\tilde\lambda + \delta) \Big( e^{\mu_2h} + \\
&  e^{\mu_3h}\Big)+\frac{\tilde{\alpha}(\tilde{\alpha}\tilde\lambda-\tilde\lambda^2+\tilde{\alpha}\delta -\delta^2)}{2w}(e^{\mu_2h}-e^{\mu_3h}) \Big]-v_{i\infty}
\end{split}
\end{equation}
\end{minipage}
}
where 

\begin{align}
    &\mu_{2,3} = -\frac{1}{2}(\delta+\tilde\lambda+\tilde\alpha) \pm w\\
    & w = \frac{1}{2}\sqrt{(\delta+\tilde\lambda+\tilde\alpha)^2-4(\tilde\lambda\delta + \tilde\alpha\delta + \tilde\lambda\tilde\alpha)}
    \label{omega}
\end{align}

that corresponds to a damped oscillator with angular frequency $w$ when the eigenvalues of the transition rate matrix become complex.

\subsubsection{Conditions for the emergence of oscillations}
\label{analytical_conditions}
We observe in simulations that oscillations emerge only when the system reaches a metastable steady state, which occur for sufficiently large spreading rate $\lambda$. This behavior contrasts with the trivial absorbing state, where the rumor does not spread through the network and both $y$ and $z$ vanish ($y, z = 0$). Analytically, we have shown that oscillations can only occur when the frequency $w$ (given by Eq. (\ref{omega})) becomes complex, that is when the argument of the square root in the expression is negative. For a random regular network with average degree $k$, this condition when $y = 0$ simplifies to:
\begin{equation}
    \delta^2+(\alpha kz_{\infty})^2-2\alpha kz_{\infty}\delta > 0
\end{equation}
From this expression, we see that when $y = 0$, $w$ is always real. Thus, oscillations are analytically ruled out in the absorbing state, and can only arise in the regime where $y>0$ at the metastable state where the rumor spreading persists. Similar to the SIS (Susceptible-Infected-Susceptible) model in epidemiology, the system exhibits a phase transition at the critical spreading rate $\lambda_c$, below which it remains in the absorbing state (rumor-free) state and above which it reaches an active metastable regime. This provides strong evidence that the appearance of oscillations is only possible above the critical threshold $\lambda_c$. 

To explore this regime analytically, we study the fixed points of the mean-field equations (Eq. (\ref{MF_eqs})) in the steady state. Fixed points correspond to the long-term behavior of the system when the time derivatives vanish. In addition to the absorbing state fixed point mentioned above, the system admits a non-trivial (metastable) fixed point. For a random regular network, where all nodes are assumed to be statistically identical ($x_i = x, y_i = y$ and $z_i = z$) and using the identity, $\sum_{i = 1}^NA_{ki}=k$ along with the normalization condition $x+y+z = 1$, we obtain the following steady-state solution:
\begin{equation}
\begin{split}
    & x_{\infty} = \frac{\alpha}{\alpha + \lambda} \\ 
    & y_{\infty} = \frac{\lambda\delta}{\alpha\lambda k + \alpha\delta + \lambda\delta} \\
    & z_{\infty} = \frac{\alpha \lambda^2k}{(\alpha + \lambda)(\alpha\lambda k + \alpha\delta + \lambda\delta)}
    \label{MF_sol_RR}
\end{split}
\end{equation}
This non-trivial fixed point characterizes the metastable regime in which a non-zero fraction of spreader and stiflers coexist, and from which oscillatory behavior may emerge depending on the parameter values.

\begin{figure}[ht]

\includegraphics[width=\linewidth]{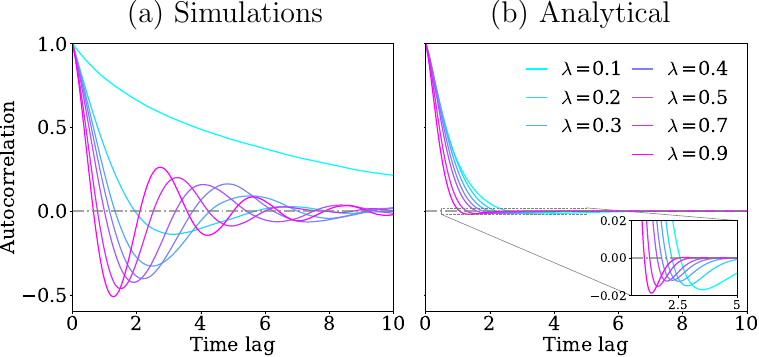}
    \caption{
       Autocorrelation as a function of the lags for different spreading rates. Plot (a) shows results from simulations, oscillating at $\lambda > \lambda_c$. Plot (b) shows the NIMFA autocorrelation as a function of the lag at different rates. Parameters: $N= 1000, \langle k\rangle = 10$, $\alpha=0.5$ and $\delta=1$.
        }
    \label{fig_autocor_sim_analytic}
\end{figure}

\begin{figure}[ht]
    \centering
    \includegraphics[width=0.5\linewidth]{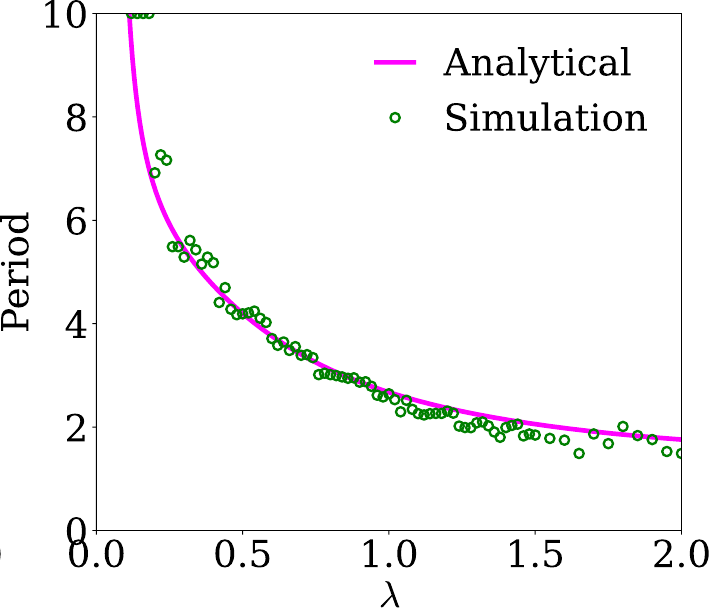}
    \caption{Period of the autocorrelation as a function of lags at different rates ($\lambda$). The analytical approximation (pink line) is computed directly from the equations. The green circles are for the simulations computed with Lomb-Scargle periodogram \cite{lomb_least-squares_1976, Scargle}. Parameters: $\langle k\rangle = 10, N = 1000$, $\alpha=0.5$ and $\delta=1$.}
    \label{fig_period_frequency}
\end{figure}

We now compare the analytical results with simulations. In Fig. \ref{fig_autocor_sim_analytic}a we show the autocorrelation of the density of spreaders obtained from simulations. In Fig. \ref{fig_autocor_sim_analytic}b, we show the NIMFA autocorrelation for a random regular network. Although the analytical results do not visibly display oscillations as clearly as in the simulations, we can still compute the oscillation period from the model equations and compare it to the period of the simulations. As shown in Fig. \ref{fig_period_frequency}, the analytically derived period agrees well with that observed in simulations. It is important to note that this agreement holds only for specific parameter values; in this case, $\alpha=0.5$ and $\delta = 1$.

\begin{figure}[h!]
\centering
\includegraphics[width=0.9\linewidth]{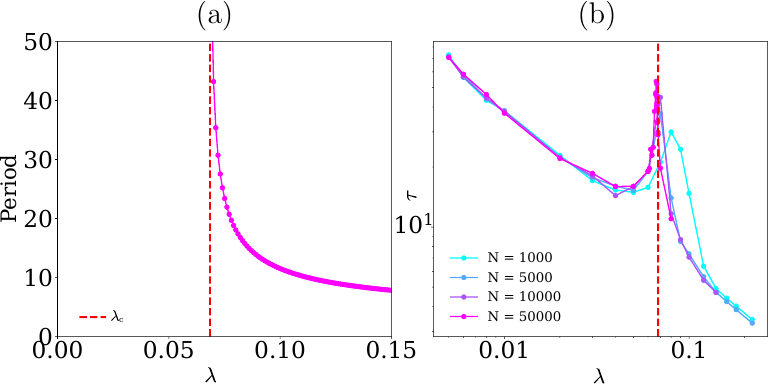}
    \caption{(a) Analytical approximation of the period of the autocorrelation as a function of the spreading rate ($\lambda$). The red dashed line indicates the lower bound of the existence of oscillations. (b) Lifetime, the red dashed line (lower bound of oscillatory behavior) is compatible with an approximation of $\lambda_c$. The parameters are $N = 1000, k = 10$, $\alpha=0.5$ and $\delta=1$.}
    \label{fig_CP}
\end{figure}

We can approximate the lower and upper bounds for the spreading rate $\lambda$ that allow oscillations by substituting the steady-state values from Eq. (\ref{MF_sol_RR}) into Eq. (\ref{omega}) using the parameters $\delta=1, \alpha=0.5$ and $k = 10$. Under these conditions, oscillations are present in the range $0.0680854 <\lambda < 6.03105$. The lower bound is compatible with an approximation of the critical point $\lambda_c$ for the same parameter set (see Fig. \ref{fig_CP}). The upper bound, however, does not align with simulations, where oscillations continue to persist beyond this value.

We also aim to understand what happens when we vary the parameters $\delta$ and $\alpha$. Since the steady-state values $y_\infty$ and $z_\infty$ inherently depend on these parameters, we will analyze the limiting behavior of Eq. (\ref{omega}) derived from the NIMFA autocorrelation as $\delta$ and $\alpha$ vary. If $y_\infty$ and $z_\infty$ are well-defined in the limit $\delta=0$, then Eq. (\ref{omega}) yields purely real eigenvalues, as the square root reduces to a perfect square. As $\delta$ increases, the equation similarly shows that the eigenvalues can remain real if $\delta$ is sufficiently large, eliminating oscillatory behavior. 

To study this behavior quantitatively, we substitute the steady-state values $y_\infty$ and $z_\infty$, obtained from the fixed point analysis Eq. (\ref{MF_sol_RR}), into Eq. (\ref{omega}). In particular, the NIMFA predicts that the critical transmissibility $\lambda_c$ increases with the forgetting rate $\delta$, indicating that faster forgetting makes spreading mode difficult. However, simulations reveal the opposite: increasing $\delta$ lowers $\lambda_c$, although the period shrinks and oscillations become noisier. This discrepancy arises because in the real dynamics, stiflers accumulate around spreaders and inhibit the transmission of the rumor; a large $\delta$ removes these nodes more quickly and thus facilitates spreading. The NIMFA does not account for such nearest-neighbor correlations and it cannot capture this mechanism. 

We also analyze the effect on varying the parameter $\alpha$, which governs how quickly spreaders become stiflers. Larger values of $\alpha$ lead to faster stifling and consequently, higher values of $\lambda_c$ are required for the rumor to persist. In this case, both the mean-field predictions and simulations show qualitatively consistent trends ($\lambda_c$ increases with $\alpha$) although the numerical values differ. A more detailed analysis can be found in Appendix \ref{conditions_osc}.

Overall, both simulations and mean-field analysis show that oscillations arise in the active metastable regime, which appears above the critical spreading rate $\lambda_c$. By examining the steady-state solutions of the mean-field equations, we determine the conditions for oscillatory behavior and identify the parameter ranges that support it. Although the NIMFA autocorrelation correctly predicts the critical point, it deviates from simulations by failing to account for local interaction effects.

\subsubsection{Case study}
\label{case_study}

High-impact events in online platforms often give rise to viral dynamics through multiple and coexisting mechanisms. In particular, bursts of attention may result from external amplification, such as mass media exposure, or from changes in the internal dynamics that enhance transmissibility. In the context of our framework, these mechanisms correspond to distinct types of virality: exogenous virality driven by external shocks and endogenous virality emerging from the system's internal dynamics. As in the synthetic scenarios considered above, real-world data typically combine both effects, making it challenging to disentangle them based solely on activity volume.

\begin{figure}[h!]
    \centering
    \includegraphics[width=0.7\linewidth]{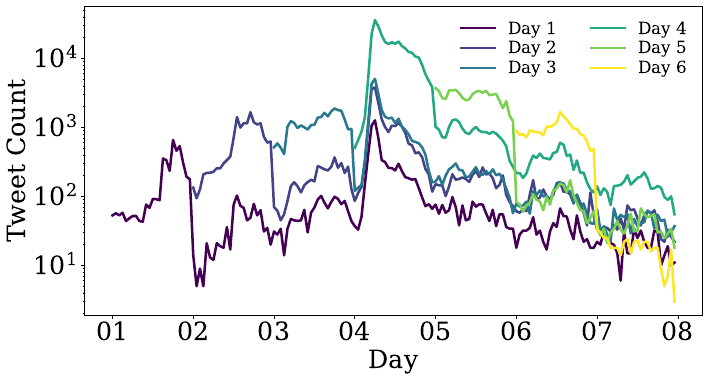}
    \caption{Tweet count time series for each group of daily new users from the Higgs boson discovery Twitter dataset.}
    \label{fig_twitter_activity}
\end{figure}

\begin{figure*}[ht!]
    \centering
    \includegraphics[width=0.9\linewidth]{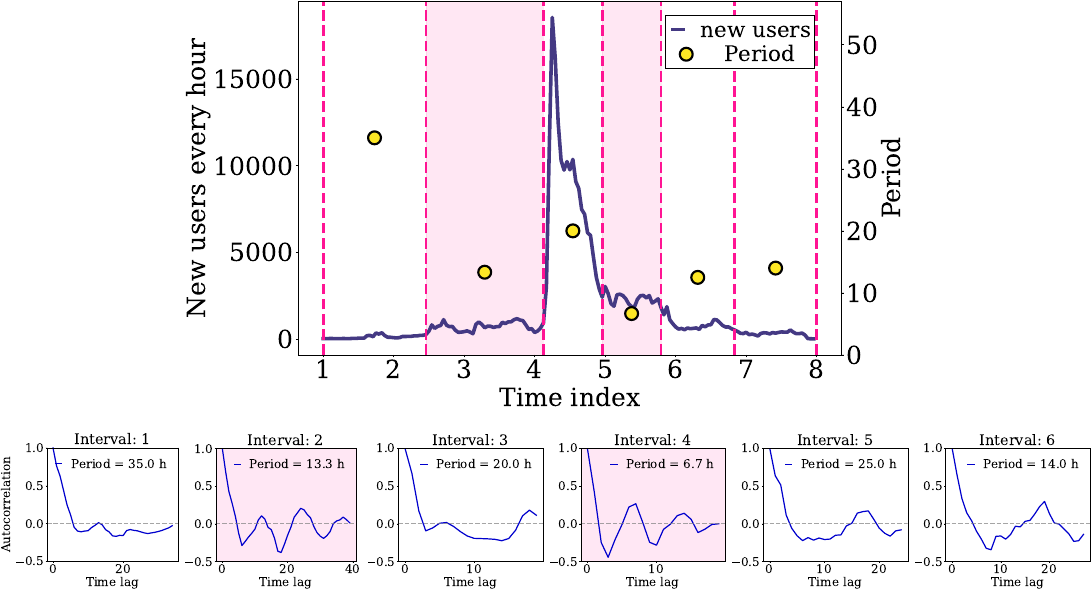}
    
    \caption{Analysis of time series of the Twitter activity surrounding the Higgs boson discovery. The main plot show the time series of new spreaders every hour, having a peak at the day 4th, containing also the statistical breakpoints. The smaller plots are the autocorrelation computed from the time series for each time series breakpoints segment. Then, the period of the autocorrelation (right axis) oscillations is plotted in circles at the main plot. The analyzed segments before and after the peak are highlighted in red.}
    \label{fig_higgs_autocor}
\end{figure*}

Here, we apply our method to a real-world dataset consisting of Twitter activity surrounding the announcement of the Higgs boson discovery, collected between July 1 and July 7, 2012 \citep{de_domenico_anatomy_2013}. This dataset provides a valuable opportunity to study the emergence of viral behavior during a high-impact event characterized by an external shock.

Our goal is to determine whether the internal dynamics of rumor transmission changed following the official announcement of the discovery on July 4th. Specifically, we aim to infer whether the observed increase in activity reflects a genuine increase in transmissibility among users, indicative of endogenous virality, or whether it was driven by an influx of new participants attracted by widespread media coverage, corresponding to exogenous virality. In other words, we aim to distinguish between an intrinsic change in the rumor spread and an increase in volume caused by external user intrusion.

We observe that users who joined the conversation in the early days continued tweeting throughout the week, and that the announcement on July 4th led to both a sharp increase in activity and a large influx of new users (see Fig. \ref{fig_twitter_activity}).

To analyze the data within the methodological framework introduced above, we segmented the new spreaders time series using a breakpoints detection method \citep{truong_selective_2020}, allowing us to isolate the system's behavior before and after the announcement. For each segment, we calculated the autocorrelation at different lags (Fig. \ref{fig_higgs_autocor}). Notably, after the peak in activity, the oscillation period in the autocorrelation signal shortens, a pattern which according to our analysis is compatible with an increase in transmissibility. This suggests that beyond the external amplification from media coverage, the internal spread of the rumor on Twitter became more intense.

From an applied perspective, this approach provides a form of nowcasting, enabling the inference of real-time shifts in the system's internal state from limited data. In practice, observing only half of an oscillation period is sufficient to estimate chages in transmissibility. Remarkably, our results align with those obtained from more computationally intensive Monte Carlo simulations on the same dataset \citep{ferraz_de_arruda_general_2018}, but our approach is significantly simpler and more efficient to implement.

\subsection{Control strategies}
\label{control_strategies}
In this final section we explore control strategies of rumor propagation. In practical settings, such as mitigating disinformation or prolonging beneficial awareness campaigns, being able to control how long a rumor persists in a network is highly valuable. 

Based on the subcritical regime results from \cite{ferraz_de_arruda_subcritical_2022}, we investigate whether the targeted placement of spreaders can modify the lifetime of a rumor, defined as the average time it takes for a finite simulation starting from a single spreader to reach the absorbing (rumor free) state.

\subsubsection{Extending lifetime}
\begin{figure}[h!]
  \centering
  \includegraphics[width=\linewidth]{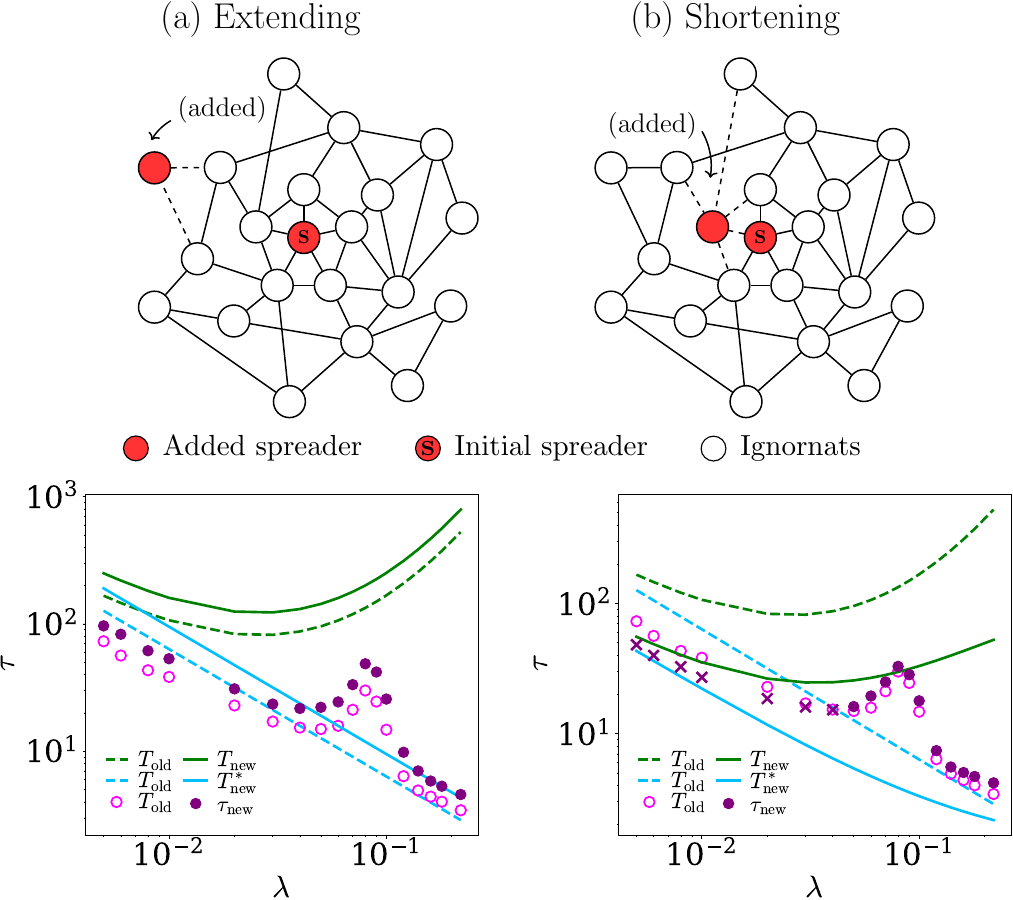}
  \caption{Network representation (top) and lifetime results (bottom) for the two intervention scenarios: \textbf{(a)} addition an external spreader far from the initial spreader, and \textbf{(b)} addition of an external spreader close to the initial spreader. The initial spreader is denoted by $s$. Filled (empty) circles indicate the lifetime with (without) intervention. Solid (dashed) lines show the corresponding analytical approximations. Crosses mark cases where intervention reduces the lifetime. Parameters: $\alpha = 0.5, \delta = 1$, $N = 1000$ and $\langle k\rangle = 10$.}
  \label{fig_rumor_control}
\end{figure}

Firstly, we consider a simple intervention: adding a new spreader at the beginning of the simulation, placed at random distance from the seed, as shown in Fig. \ref{fig_rumor_control}A. This minimal change leads to a substantial increase in the average lifetime of the rumor. In the subcritical regime, where spreading is scarce and localized, each spreader initiates an independent cascade, and the system only reaches the absorbing state once both cascades have terminated. Therefore, the total lifetime can be approximated as the maximum time of two independent exponentials:
\begin{equation}
    T_{\text{independent spreaders}} \approx \frac{3}{2}T_{\text{single spreader}} 
\end{equation}

This behavior is supported by the analytical lifetime approximation developed in \cite{ferraz_de_arruda_subcritical_2022}, which considers the minimal set of transitions required to reach the absorbing state. For a single initial spreader at node $i$, the shortest extinction sequence is:

\begin{enumerate}[leftmargin=*, labelsep=0.5em, itemsep=0pt]
    \item Node $i$ spreads the rumor to a neighbor $j$, $\left\langle {\tau }_{1}\right\rangle$ 
    \item \label{step_stifler1}One of the two nodes ($i$ or $j$) becomes a stifler, $\left\langle {\tau }_{2}\right\rangle$
    \item \label{step_stifler2}The second node becomes a stifler, $\left\langle {\tau }_{3}\right\rangle$ 
    \item A stifler forgets the rumor and becomes ignorant, $\left\langle {\tau }_{4}\right\rangle$
    \item The remaining node also forgets the rumor, $\left\langle {\tau }_{5}\right\rangle$
\end{enumerate}

Since transitions are governed by the continuous time Markov chain, the total lifetime can be approximated by counting the number of times the process fails to reach the absorbing state plus the time it succeed \cite{ferraz_de_arruda_subcritical_2022}.
\begin{equation}
    T=\mathop{\sum }\limits_{i=1}^{\infty }i\left\langle {\tau }_{1}\right\rangle {\left(1-{P}_{{{{{\rm{abs}}}}}}\right)}^{i}+{P}_{{{{{\rm{abs}}}}}}\left\langle {\tau }_{1\to 5}\right\rangle \approx \left\langle {T}_{{{{{\rm{abs}}}}}}\right\rangle 
\end{equation}

\begin{figure}[h]
\centering
    \includegraphics[width=0.55\linewidth]{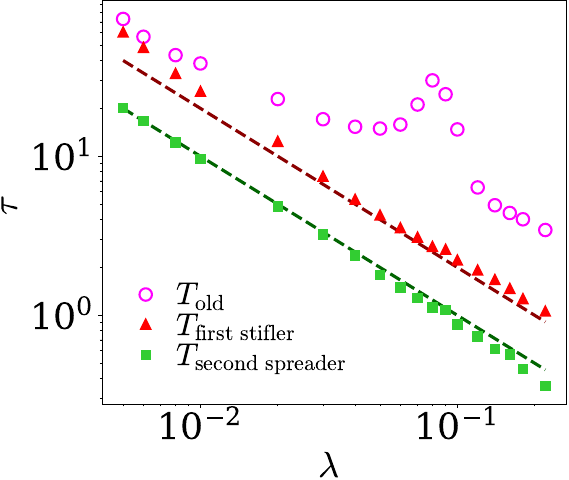}
    \caption{Average temporal observables of the process starting from a single spreader. Circles indicate the lifetime. Triangles denote the average time of appearance of the first stifler, and squares the average time of appearance of the second spreader. Dashed lines represent the corresponding analytical approximations. Parameters: $\alpha = 0.5, \delta = 1$, $N = 1000$ and $\langle k\rangle = 10$.}
    \label{fig_key_events}
\end{figure}

As explained in \cite{ferraz_de_arruda_subcritical_2022}, in the subcritical regime the time that dominates is that of spreading events. Knowing this we can know with precision the time it will take for the second spreader and the first stifler to appear, as seen in Fig. \ref{fig_key_events}. In contrast with the previous results, we note that this simple analytical approximation of the phenomenon fits perfectly with the simulations. Although we have not designed any intervention for this case, we believe that it may be useful for further research. On average, the time of the appearance is:
\begin{equation}
    \langle \tau_{{2_{nd} \ \text{spreader}}} \rangle = \frac{1}{k\lambda} \ \text{and} \ \langle \tau_{1_{st}\ \text{stifler}} \rangle = \frac{2}{k\lambda}
\end{equation}

\subsubsection{Shortening lifetime}
\label{reducing_lifetime}
While extending the lifetime of a rumor may be desirable in cases like public awareness campaigns, there are many contexts (such as combating disinformation) where the goal is the opposite: to shorten the rumor's duration and minimize its impact. We now explore whether the lifetime of a rumor can be reduced through targeted interventions.

To this end, we introduce a second spreader in close proximity to the seed node. Specifically, we place the new spreader in a neighboring node before the first spreading event occurs. In the subcritical regime, where spreading is already inefficient, this small change has a significant effect. Because the two spreaders are so close, they are likely to interact early. This increases the likelihood of early extinction, effectively bypassing the long transient dynamics characteristic of the subcritical regime conditions.

By placing two spreaders close together, we effectively increase the chance that steps (\ref{step_stifler1}) and (\ref{step_stifler2}) from the events sequence occur earlier, leading to early extinction of the rumor. Simulations confirm that this simple intervention can significantly shorten the expected lifetime in the subcritical regime (see Fig. \ref{fig_rumor_control}B). Analytically, we approximate it by adding a probability that the two spreaders interact and on average the process finish earlier. The full derivation can be found in Appendix \ref{lifetime_approx}. The time to reach the absorbing state for $\lambda \ll 1$ and $\lambda \ll \alpha,\delta$ regime can be approximated as:

\normalsize

\begin{equation}
\begin{split}
    T^{\ast} = \frac{1}{{2(\langle k\rangle }-1)\lambda} \cdot \frac{1- \left( \frac{1}{2\langle k\rangle} + \frac{\alpha}{\alpha + \delta}\right)}{\left( \frac{1}{2\langle k\rangle} + \frac{\alpha}{\alpha + \delta}\right)^2} + \\
    \left( \frac{1}{2\langle k\rangle} + \frac{\alpha}{\alpha + \delta}\right)\cdot  \left( \frac{1}{\alpha + \delta} + \frac{1}{2\alpha} + \frac{3}{2\delta}\right)
\end{split}
\end{equation}

These results highlight how proximity-based interventions can be strategically employed to modify rumor lifetime, based on a lightweight and scalable method.

\section{Discussion}
\label{discussion}

This study advances our understanding of rumor spreading in complex networks and provides practical tools that are applicable to real-world social systems. We show that early warning signals, such as autocorrelation, can be used not only to detect the onset of viral behavior in rumors but also to infer shifts in the underlying transmission dynamics.
Building on this, we propose strategies to either extend or shorten the lifetime of a rumor, enabling more effective amplification or suppression depending on the application.

We explored whether a sudden burst in activity caused by external events, such as media broadcasts or astroturfing campaigns, can be distinguished from genuine increases in internal transmissibility. By analyzing the temporal evolution of the lag-1 autocorrelation, computed as a rolling average over time, we find that different spreading mechanisms leave distinct statistical signatures. External spreaders introduced near active nodes lead to an increase in autocorrelation, potentially mimicking the effect of higher transmissibility, and the temporal pattern depends on whether these spreaders remain permanently active or follow the same dynamics as others. In the latter case, the autocorrelation decays more slowly after the perturbation.

This results are especially of interest from a practical perspective. In fact, previously proposed astroturfing detection methods rely on the analysis of interaction networks, linguistic content, and user metadata \cite{keller_political_2020,ratkiewicz2011truthy}, being computationally and data intensive, our results allow early detection methods based on the statistical analysis of the new spreader time series only.

While temporal lag-1 autocorrelation alone may be insufficient to distinguish these scenarios, analyzing the autocorrelation as a function of the lag reveals a crucial signal: only internally driven growth produces oscillatory patterns across lags. These oscillations arise when the system enters a metastable regime above the critical transmissibility threshold $\lambda_c$. Using the NIMFA, we derived an analytical expression for this behavior as a damped oscillator when the eigenvalues of the infinitesimal generator matrix are complex. Although oscillations are strongly damped in the analytical model, the predicted period aligns closely with simulations, allowing analytical estimation of the critical point and establishing autocorrelation as a robust indicator for identifying spreading mechanisms.

The emergence of oscillations connects naturally to non-equilibrium statistical mechanics. It is well established that oscillations cannot occur in equilibrium steady-state systems. Oscillations require the system to be driven away from equilibrium, which in stochastic terms corresponds to a non-zero net flux (a measure of non-equilibrium strength) that breaks the detailed balance. For systems with three or more states, theory predicts that at a critical flux value a phase transition can occur, producing oscillatory overshoots \cite{overshoot, jia_nonequilibrium_2016}. From a Markov chain perspective, this transition corresponds to the generator matrix acquiring complex conjugate eigenvalues with negative real parts.

Returning to our model, we observe notable discrepancies between the NIMFA predictions and simulations when varying parameters such as the forgetting rate $\delta$. While the NIMFA suggests that higher $\delta$ hinders spreading, simulations reveal the opposite. This mismatch stems from the NIMFA's neglect of local correlations, such as stifler clustering around spreaders, which can block further transmission. A higher forgetting rate reduces these local obstructions, indirectly facilitating spread.

To validate our approach, we applied it to Twitter activity surrounding the July 2012 announcement of the Higgs boson discovery. Our aim was to determine whether the post-announcement spike was driven solely by media amplification or also by an increase in internal transmissibility. Measuring autocorrelation at multiple lags before and after the peak, we found a higher frequency of oscillations afterward, consistent with an internal transmissibility increase. This result demonstrates the method's applicability to real-world data.

We further demonstrate that targeted interventions can substantially alter a rumor's duration. These interventions involve introducing external spreaders at specific network locations. A rumor lasts longer when the new spreader is placed far from existing ones, as spatial separation leads to the formation of independent spreading processes that interfere less with each other. Conversely, placing a spreader near an existing one shortens the lifetime in the subcritical regime. This accelerates the transitions from spreader to stifler states where the rumor is slow and reaches few nodes. As part of this framework, we also estimate the timing of key events, such as the appearance of a second spreader or a stifler. Interestingly, extending a rumor's lifetime is generally easier than shortening it, an insight that aligns with intuitive expectations. These results extend and complement previous ones \cite{belen_impulsive_2005} by elucidating the role of the modified expected lifetime of the rumor when new spreaders are introduced in the populations and their locations.

In summary, we show that autocorrelation analysis across multiple lags can reliably distinguish internally driven spreading from externally introduced activity. Oscillatory patterns are unique to endogenous growth and can be predicted analytically, with the mean-field approximation accurately capturing both the critical threshold and the oscillation period for certain parameter ranges, even though its performance declines when local correlations become dominant. Also, we demonstrated that the lifetime of a rumor can be controlled through targeted placement of spreaders within a network.

These findings contribute not only to the theoretical modeling of rumor dynamics but also have broader implications for online communication environments. By identifying how rumor volume and visibility can be artificially amplified, our findings help explain how “false volume” can be created in online platforms. In doing so, we highlight mechanisms that are often employed by powerful actors, and facilitated by large technology platforms, to manipulate online discourse using fake accounts and coordinated activity. This challenges the notion of social networks as neutral spaces for information exchange and underscores the importance of developing tools to detect and mitigate such manipulation.

\begin{acknowledgments}
This work was financially supported by the Catalan Government through the funding grant ACCIÓ-Eurecat (Project TRAÇA2023 - CLINT). E.R. is a fellow of Eurecat’s “Vicente Lopez” PhD grant program, E.C. acknowledges support from the Spanish grants PID2021-128005NB-C22, funded by MCIN/AEL 10.13039/501100011033, and from Generalitat de Catalunya under project 2021-SGR-00856.
\end{acknowledgments}

\section*{Author contributions}
E.R., J.V. and E.C. conceptualized the study, E.R. performed data analyses, wrote the original draft and all authors critically discussed the results, revised the paper and approved the final manuscript.

\section*{Data Availability}
Some of the data and code that support the findings of the article are openly available. The simulation code used to reproduce the main results is available at \cite{rifa2026ruscs}. The Twitter dataset analyzed is available at \cite{de_domenico_anatomy_2013}. Additional simulation data and extended analysis scripts cannot be made publicly available due to size and and computational reproducibility constraints but are available from the authors upon reasonable request.

\appendix

\section{Methods}
\label{methods}
\subsection{Networks}
Synthetic networks were generated in Python using NetworkX’s \citep{hagberg_exploring_2008} functions: \texttt{random\_regular\_graph(d,n)} for the random regular networks \citep{steger_generating_1999}, \texttt{powerlaw\_cluster\_graph(n,m,p)} for the clustered scale‐free \citep{holme_growing_2002} and \texttt{erdos\_renyi\_graph(n,p\_er)} for the Erdős-Rényi graph \citep{Erdos}. 

\subsection{Monte Carlo Simulations}
We use an optimized Gillespie algorithm \citep{Gutjahr_2021} to simulate the exact stochastic trajectories of the continuous-time Markov process \citep{gillespie_exact_1977}. 

In order to collect significant statistics in the subcritical regime where the system would rapidly reach the absorbing state and to approximate the steady-state behavior, we apply the quasi-stationary (QS) method \citep{PhysRevE.71.016129}. With it, we prevent the system from being absorbed by reintroducing a saved active configuration when the absorption state is reached.
\subsection{Lifetime}
\label{methods_lifetime}
To compute the lifetime of finite realizations, each run is initialized with a single randomly chosen spreader and the rest of the population as ignorants. The simulation proceeds until the system reaches an absorbing state \citep{boguna_nature_2013}. 

\subsection{Early warning signals indicators}
\subsubsection{Mutual Information}
The mutual information (MI) quantifies the amount of information that one random variable contains about another, measuring the statistical dependence between them and capturing linear and non-linear relationships. It is defined as \cite{george_early_2023}:
\begin{equation}
    M_{X,Y} = \sum_{i,j} p_{X,Y}(i,j)\log\Biggr(\frac{p_{X,Y}(i,j)}{p_X(i)p_Y(j)}\Biggr)
\end{equation}
Where $p_{X,Y}(i,j)$ is the joint probability distribution of variables $X$ and $Y$, and $p_X(i)$ and $p_Y(j)$ are their corresponding marginal distributions. We have used \texttt{mutual\_info\_score} from Python library \texttt{scikit-learn} \cite{scikit-learn}. Specifically, we have computed the mutual information over symbolic time series on a rolling time window of 50 steps picking up two nodes at a random distance. Then we have computed the average of 1000 realizations for each time step.

\subsubsection{Autocorrelation}
Autocorrelation measures the correlation of a signal with a delayed copy of itself. To compute it we have used function \texttt{statsmodels.tsa.stattools.acf} from Python library \texttt{statsmodels} \cite{seabold2010statsmodels}. Similar to the mutual information we have compute it with a rolling window of 50 steps, from the spreaders density time series, then averaged over 1000 realizations.

\section{Autocorrelation for different network structures}
\label{oscillations_structures}
\begin{figure}[h!]
    \centering
    \includegraphics[width=0.6\linewidth]{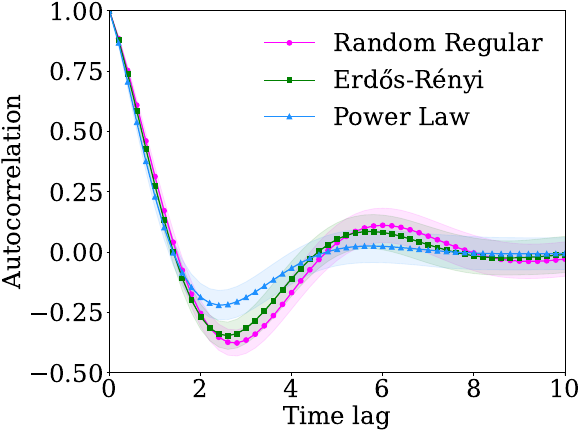}
    \caption{Autocorrelation of spreaders density as a function of lag for a high spreading rate $\lambda = 0.3$ for an Erdős–Rényi and a power-law network ($\gamma \approx 3$) with $\langle k\rangle \approx10$. The period of the oscillations is similar, the Erdős–Rényi has a similar behavior with the random regular.}
    \label{fig_lags_net_struct}
\end{figure}

In this section, we analyze the autocorrelation of the density of spreaders when the transmissibility rate $\lambda$ is high, across different network structures. We generated a clustered power-law network (degree exponent $\approx$ 3) whose maximum degree was 196 and whose average degree was 9.932, and an Erdős–Rényi random graph with a maximum degree of 21 and an average degree of 9.972. In Fig. \ref{fig_lags_net_struct}, we plotted the autocorrelation as a function of lag and we observe clear oscillations.

\section{The NIMFA autocorrelation in the steady state}
\label{analytica_autocor}
\subsection{Analytical derivation}
We derive the autocorrelation function for the spreading state of a node in the steady state of the Maki-Thompson modified model using the $N$-intertwined mean-field approximation (NIMFA) \citep{Van_Mieghem_2014,liu_autocorrelation}.

The autocorrelation of the spreading state $V_j(t)\in \{0,1\}$ of a node $j$ between times $s$ and $t$ is defined as:
\begin{align}
    R_{j}(s,t) \triangleq \frac{E[V_{j}(s)V_{j}(t)]-E[V_{j}(s)]E[V_{j}(t)]}{\sqrt{Var[V_{j}(s)]Var[V_{j}(t)]}}
\end{align}
In the steady state ($t\rightarrow\infty$), we denote the time difference as lag $h$, and write $R_{j\infty}(h)\equiv R_j(t,t+h)$.

Using the NIMFA, we can define effective transmission and stifling rates. For node $i$, the steady-state effective rates are:

\begin{equation}
    \tilde{\lambda}_{i} \equiv \lambda \sum^N_{k = 1}A_{ki}y_{i\infty} \quad \text{ and } \quad \tilde{\alpha}_{i} \equiv \alpha \sum^N_{k = 1}A_{ki}(y_{k\infty} + z_{k\infty})
    \label{effective_rates_app}
\end{equation}
These rates define the mean-field evolution of the spreading probability $v_i(t)$, approximated as:
\begin{equation}
    \frac{dv_i(t)}{dt} = \left[1-v_i(t)-\frac{1}{\delta}v_i(t)\tilde{\alpha}\right]\tilde{\lambda}- v_i(t)\tilde\alpha.
\end{equation}

The non-trivial steady-state spreading probability is:
\begin{equation}
    v_{i\infty} =\frac{\tilde\lambda}{\tilde\lambda(1+\frac{\tilde\alpha}{\delta})+\tilde\alpha}
\end{equation}

Additionally, one can bound this by assuming a maximum neighborhood degree $d_i$:
\begin{equation}
    0 \leq v_{i\infty} \leq \frac{\lambda}{\lambda (1+\frac{\alpha d_i}{\delta})+\alpha}
\end{equation}

We now construct the infinitesimal generator $Q_{i}$ for the Markov process governing the state transitions of node $i$
\begin{equation}
    Q_{i\infty} = \begin{pmatrix}
        -\tilde{\lambda}_{i} & \tilde{\lambda}_{i} & 0 \\
        0 & -\tilde{\alpha}_{i} & \tilde{\alpha}_{i} \\
        \delta & 0 &-\delta
    \end{pmatrix}
\end{equation}
this defines the transitions between ignorant, spreader and stifler states. The transition matrix at lag $h$ is:
\begin{equation}
    P_{i\infty}(h) = e^{Q_{i\infty}h} = U_ie^{\Lambda_i}U^{-1}_{i}.
\end{equation}
where $U_i$ and $\Lambda_i$ are the eigenvector matrix and the diagonal eigenvalue matrix of $Q_{i\infty}h$, respectively. The eigenvalues of $Q_{i\infty}$ are:
 \begin{equation}
 \label{eigenvaules}
     \mu_{2,3} = -\frac{1}{2}(\delta+\tilde\lambda+\tilde\alpha) \pm \frac{1}{2}\sqrt{(\delta+\tilde\lambda+\tilde\alpha)^2-4(\tilde\lambda\delta + \tilde\alpha\delta + \tilde\lambda\tilde\alpha)}
 \end{equation}
Let us define:
\begin{equation}
\begin{split}
 &a = -\frac{1}{2}(\delta+\tilde\lambda+\tilde\alpha)\\
 &w = \frac{1}{2}\sqrt{(\delta+\tilde\lambda+\tilde\alpha)^2-4(\tilde\lambda\delta + \tilde\alpha\delta + \tilde\lambda\tilde\alpha)}  
\end{split}
\end{equation}

We focus on the transition probability of remaining spreader:
\begin{equation}
    P_{i\infty}(h)_{2,2} = Pr[V_{i\infty}(t+h) = 1 | V_{i\infty}(t) = 1]
\end{equation}
After simplifying, we get expression:

\begin{equation}
\begin{split}
P_{i\infty}(h)_{2,2} = &\frac{1}{\tilde{\alpha}\tilde\lambda+\tilde{\alpha}\delta +\tilde\lambda\delta} \Big[
        \delta\tilde\lambda + \frac{\tilde{\alpha}}{2}(\tilde\lambda + \delta) \Big( e^{\mu_2h} + e^{\mu_3h}\Big) \\
    &+\frac{\tilde{\alpha}(\tilde{\alpha}\tilde\lambda-\tilde\lambda^2+\tilde{\alpha}\delta -\delta^2)}{2w}(e^{\mu_2h}-e^{\mu_3h}) \Big]
\end{split}
\end{equation}

Since in the steady state $E[V_{i\infty}(t)]=v_{i\infty}$, the autocorrelation is:

\begin{equation}
    R_{i\infty}(h)
    = \frac{P_{i\infty}(h)_{2,2}-v_{j\infty}}{1-v_{i\infty}}
\end{equation}
Substituting $P_{i\infty}(h)_{2,2}$, we obtain the expression for the autocorrelation:
\begin{equation}
\begin{split}
    & R_{i\infty}(h)
    = \\ 
    & \frac{1}{1-v_{i\infty}} \Biggr[ \frac{1}{\tilde{\alpha}\tilde\lambda+\tilde{\alpha}\delta +\tilde\lambda\delta} \Big[
        \delta\tilde\lambda + \frac{\tilde{\alpha}}{2}(\tilde\lambda + \delta) \Big( e^{\mu_2h} + e^{\mu_3h}\Big)\\
    &+\frac{\tilde{\alpha}(\tilde{\alpha}\tilde\lambda-\tilde\lambda^2+\tilde{\alpha}\delta -\delta^2)}{2w}(e^{\mu_2h}-e^{\mu_3h}) \Big]-v_{i\infty}\Biggr]
\end{split}
\end{equation}
When the eigenvalues ($\mu_{2,3}$) are complex, we get the closed-form expression:

\resizebox{\columnwidth}{!}{
\begingroup
\thinmuskip=0mu
\medmuskip=0mu
\thickmuskip=0mu
\everymath{\displaystyle}
{
\footnotesize
$
  R_i = \frac{1}{1-v_{\infty}}\left[\frac{1}{\tilde{\alpha}\tilde\lambda+\tilde{\alpha}\delta +\tilde\lambda\delta} \left[
      \delta\tilde\lambda + Ae^{ah}\left(\cos{wh} - \phi \right) \right] - v_{\infty} \right]
$
}
\endgroup
}

where 

{
\small
\begin{align}
    &A = \sqrt{{(\tilde{\alpha}(\tilde\lambda + \delta))}^2 + {\left(\frac{\tilde{\alpha}(\tilde{\alpha}\tilde\lambda-\tilde\lambda^2+\tilde{\alpha}\delta -\delta^2)}{w}\right)}^2}\\
    &\phi=\arctan{\left(\frac{\tilde{\alpha}(\tilde{\alpha}\tilde\lambda-\tilde\lambda^2+\tilde{\alpha}\delta -\delta^2)}{w\tilde{\alpha}(\tilde\lambda + \delta)}\right)}
\end{align}
}
that corresponds to a damped oscillator.

\subsection{Conditions for the emergence of oscillations}
\label{conditions_osc}

The autocorrelation shows oscillations when the eigenvalues of the infinitesimal generator are complex. We can get a full expression if we substitute the steady-state values from Eq. (\ref{MF_sol_RR}) into Eq. (\ref{omega}):
{
\footnotesize
\begin{equation}
\label{period_eq}
\begin{split}
&\omega^2 = {\left(\delta+\frac{\lambda^2k\delta}{\alpha\lambda k + \alpha\delta + \lambda\delta}+\frac{\alpha k \lambda}{\alpha + \lambda}\right)}^2 \\
&-4\left( \frac{\lambda^2k\delta}{\alpha\lambda k + \alpha\delta + \lambda\delta}\delta+\frac{\alpha k \lambda}{\alpha + \lambda}\delta + \frac{\alpha k \lambda}{\alpha + \lambda}\frac{\lambda^2k\delta}{\alpha\lambda k + \alpha\delta + \lambda\delta}\right)
\end{split}
\end{equation}
}
The critical transmissibility $\lambda_c$ can be found with the following expression:
{
\footnotesize
\begin{equation}
\begin{split}
&\delta^2+\left(\frac{\lambda^2k\delta}{\alpha\lambda k + \alpha\delta + \lambda\delta}\right)^2+\left(\frac{\alpha k \lambda}{\alpha + \lambda}\right)^2 =\\
&-2\left( \frac{\lambda^2k\delta}{\alpha\lambda k + \alpha\delta + \lambda\delta}\delta+\frac{\alpha k \lambda}{\alpha + \lambda}\delta + \frac{\alpha k \lambda}{\alpha + \lambda}\frac{\lambda^2k\delta}{\alpha\lambda k + \alpha\delta + \lambda\delta}\right)
\end{split}
\end{equation}
}
Since we can't get a closed expression for $\lambda_c$, we compute the limits from when the Eq. \ref{omega} is complex. For the case $\lambda \ll \delta, \alpha$:
{\normalsize
\begin{equation}
    w^2 \underset{\lambda \ll \delta,\alpha}{\approx} \delta - 2k\lambda\delta + \lambda^2\left( k^2-2\frac{k}{\alpha}\delta \right) + \mathcal{O}(\lambda^3) < 0
\end{equation}
}
the minimum value for $\lambda$ at second order is:
\begin{equation}
    \lambda^* = \frac{2k\delta-\sqrt{\frac{8k\delta^3}{\alpha}}}{2(k^2-\frac{2k\delta}{\alpha})}
\end{equation}

\begin{figure}[h!]
\centering
\includegraphics[width = \linewidth]{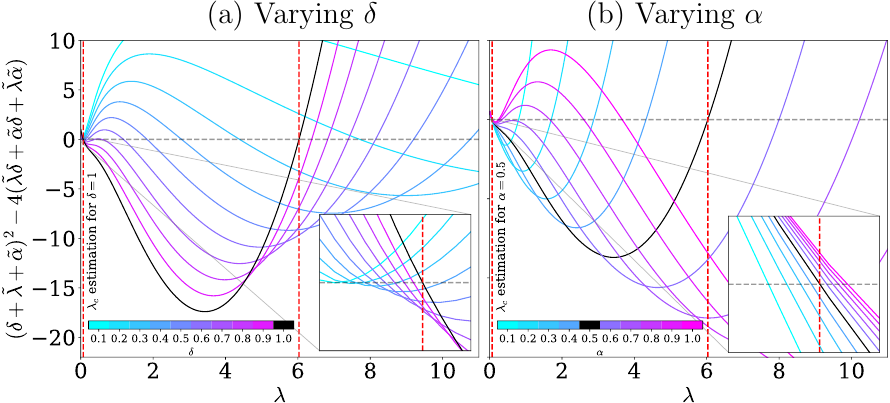}
\caption{(a) Comparison of the condition for the oscillations, varying $\delta$ from $0.1$ to $1$. When the expression is negative, the eigenvalue is complex and the oscillations appear. In black, the results with $\delta = 1$ (b) Comparison of the condition for the oscillations, varying $\alpha$ from $0.1$ to $1$. When the expression is negative, the eigenvalue is complex and the oscillations appear. In black, the results with $\alpha = 0.5$}
    \label{fig_comparison_MF}
\end{figure}

From this we can see that $\lambda_c$ will grow as $\delta$ becomes larger. In the same way, $\lambda_c$ will grow as $\alpha$ becomes larger. We can observe this in Fig. \ref{fig_comparison_MF}, where we plot the condition that make the oscillations appear as a function of $\lambda$, for different $\delta$ (Fig. \ref{fig_comparison_MF}a) and for different $\alpha$ (Fig. \ref{fig_comparison_MF}b). The oscillations start to appear when the function is negative, so $\lambda_c$ is the first time the function crosses 0.

\begin{figure}[h]
\centering
\includegraphics[width=0.9\linewidth]{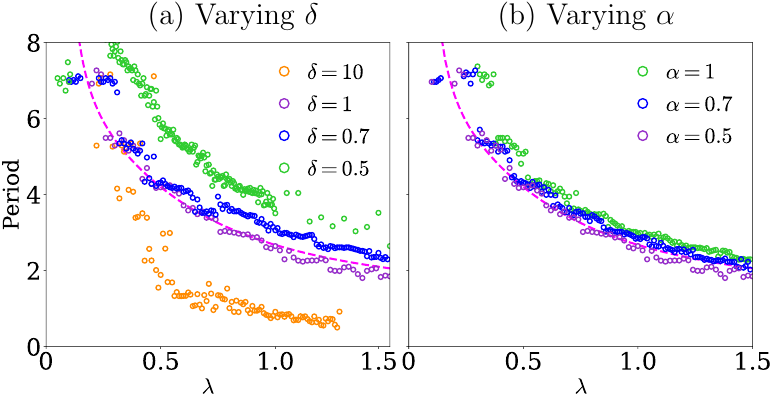}
    \caption{Average lifetime as a function of rates $\lambda$: (a) for different values of $\delta \in \{0.5, 0.7, 1\}$, with $\alpha=0.5$, and (b) for different values of $\alpha \in \{0.5, 0.7, 1\}$, with $\delta=1$. The horizontal lines indicate the approximate values of $\lambda_c$, determined from the divergence point of the lifetime.}
\label{lifeitme_delta_alpha}
\end{figure}

\begin{figure}[h] 
\centering
\includegraphics[width=0.9\linewidth]{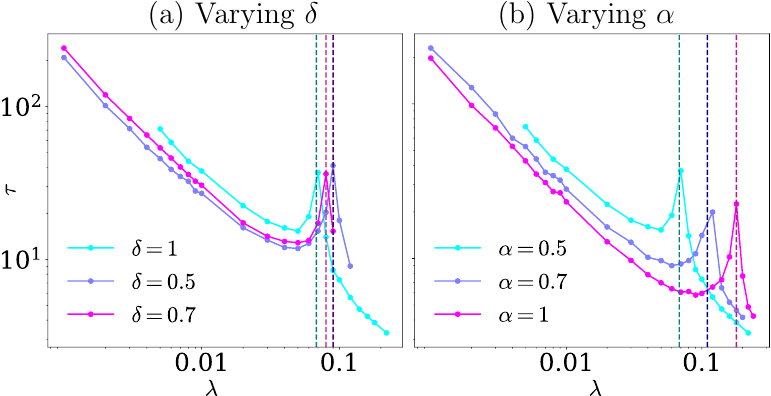}
    \caption{Period of the autocorrelation as a function of rates $\lambda$: (a) for different values of $\delta \in \{0.5, 0.7, 1, 10\}$, with $\alpha=0.5$, and (b) for different values of $\alpha \in \{0.5, 0.7, 1\}$, with $\delta=1$. The pink dashed line shows the analytical result for $\delta=1$, $\alpha=0.5$, which matches  the simulations. For other combinations of parameters, the analytical approximation do not coincide with simulations.}
\label{periods_varying_params}
\end{figure}

However, if we compare these results with the simulations, we observe important differences. We know where is $\lambda_c$ because it is when the lifetime diverges. In Fig. \ref{lifeitme_delta_alpha}a we plotted the lifetime as a function of rate $\lambda$ for different values of $\delta$. We see that the value of $\lambda_c$ decreases as the parameter $\delta$ increases. This is in contradiction with the NIMFA results. For parameter $\alpha$ Fig. \ref{lifeitme_delta_alpha}b, we see that the result matches qualitatively the behavior for the second order approximation: as $\alpha$ increases, the $\lambda_c$ increases.
Finally, we plot the period of the osculations for different parameter values $\alpha$ and $\delta$. In Fig. \ref{periods_varying_params}a we plotted the oscillations period of the simulations for different values of $\delta$. We see that as $\delta$ increases, the oscillations period decreases, and therefore, it is compatible that oscillations start to appear earlier for larger $\delta$, as we can know by looking at the lifetime. In Fig. \ref{periods_varying_params}b we can see that the period of the oscillations decreases as the parameter $\alpha$ decreases, maing also compatible the lifetime results.

\section{Approximation of the lifetime for the intervention}
\label{lifetime_approx}
The lifetime can be approximated as \citep{ferraz_de_arruda_subcritical_2022}:
assuming that node $i$ is a spreader, the shortest sequence of events that will drive the process to the absorbing state consists of: (1) node $i$ spreads the information to node $j$, (2) either $i$ or $j$ turn into stifler, (3) the other node turns into a stifler, (4) either $i$ or $j$ turn into ignorant, and (5) the other node turns into an ignorant. Since our process is a continuous Markov chain, only one event occurs at a time, the respective probabilities $P$ and expected times $\left\langle {\tau }\right\rangle$ of these events are:
{

\begin{align}
    &{P}_{1}=1\,\,\,\left\langle {\tau }_{1}\right\rangle =\frac{1}{{\langle k\rangle }\lambda } \\
    &{P}_{2}=\frac{2\alpha }{2\alpha +2\lambda ({\langle k\rangle }-1)}\,\,\,\left\langle {\tau }_{2}\right\rangle =\frac{1}{2\alpha +2\lambda ({\langle k\rangle }-1)}\\
    &{P}_{3}=\frac{\alpha }{\alpha +\lambda ({\langle k\rangle }-1)+\delta }\,\,\left\langle {\tau }_{3}\right\rangle =\frac{1}{\alpha +\lambda ({\langle k\rangle }-1)+ \delta}\\
    &{P}_{4}={P}_{5}=1\,\,\,\,\,\,\,\,\,\,2\left\langle {\tau }_{4}\right\rangle =\left\langle {\tau }_{5}\right\rangle =\frac{1}{\delta }.
\end{align}
}
The probability and the expected time for reaching the absorbing state through this chain are:
\begin{equation}
    {P}_{{{{{\rm{abs}}}}}}=\frac{{\alpha }^{2}}{(\alpha +({\langle k\rangle }-1)\lambda )(\alpha +\delta +({\langle k\rangle }-1)\lambda )}
\end{equation}
\begin{equation}
    \langle {\tau }_{1\to 5}\rangle =\frac{1}{{\langle k\rangle }\lambda }+\frac{3}{2\delta }+\frac{3}{2\alpha +2\lambda ({\langle k\rangle }-1)}
\end{equation}
In the $\lambda \ll \delta$ and $\lambda \ll \alpha$ regime, $\left\langle {\tau }_{1}\right\rangle$ is the largest time:
\begin{equation}
    T=\mathop{\sum }\limits_{i=1}^{\infty }i\left\langle {\tau }_{1}\right\rangle {\left(1-{P}_{{{{{\rm{abs}}}}}}\right)}^{i}+{P}_{{{{{\rm{abs}}}}}}\left\langle {\tau }_{1\to 5}\right\rangle \approx \left\langle {T}_{{{{{\rm{abs}}}}}}\right\rangle ,
\end{equation}

For the case the lifetime is reduced (see Section \ref{reducing_lifetime}) in the subcritical regime, the equations are:
\begin{equation}
    T= \left\langle {\tau }_{1}\right\rangle \frac{1-{P}_{{{{{\rm{abs}}}}}}}{({P}_{{{{{\rm{abs}}}}}})^2} +{P}_{{{{{\rm{dir}}}}}}\left\langle {\tau }_{2\to 4}\right\rangle +{P}_{{{{{\rm{ind}}}}}}\left\langle {\tau }_{1\to 4}\right\rangle 
\end{equation}
where
\begin{equation}
    {P}_{{{{{\rm{abs}}}}}}={P}_{{{{{\rm{dir}}}}}} + {P}_{{{{{\rm{ind}}}}}}
\end{equation}
\begin{equation}
    {P}_{{{{{\rm{ind}}}}}}=\frac{{\alpha }^{2}}{(\alpha +({\langle k\rangle }-2)\lambda )(\alpha +\delta +({\langle k\rangle }-2)\lambda )}
\end{equation}
\begin{equation}
    {P}_{{{{{\rm{dir}}}}}}=\frac{1}{2(\langle k\rangle-1)}\frac{\alpha}{\lambda+\alpha}
\end{equation}
\begin{equation}
    \left\langle {\tau }_{1}\right\rangle =\frac{1}{{(\langle k\rangle }-1)\lambda }
\end{equation}
\begin{equation}
    \left\langle {\tau }_{2\rightarrow4}\right\rangle = \frac{1}{\alpha +\delta +({\langle k\rangle }-2)\lambda }+\frac{1}{2\alpha +2({\langle k\rangle }-2)\lambda }+\frac{3}{2\delta }
\end{equation}

Finally,
\begin{widetext}
{
\footnotesize

\begin{equation}
\begin{split}
    T= \frac{1}{{2(\langle k\rangle }-1)\lambda } &\cdot \frac{1-\frac{1}{2(\langle k\rangle-1)}\frac{\alpha}{\lambda+\alpha} - \frac{{\alpha }^{2}}{(\alpha +({\langle k\rangle }-2)\lambda )(\alpha +\delta +({\langle k\rangle }-2)\lambda)}}{(\frac{1}{2(\langle k\rangle-1)}\frac{\alpha}{\lambda+\alpha} + \frac{{\alpha }^{2}}{(\alpha +({\langle k\rangle }-2)\lambda )(\alpha +\delta +({\langle k\rangle }-2)\lambda)})^2} 
    +\left(\frac{1}{2\langle k\rangle}\frac{\alpha}{\lambda+\alpha} \right) \cdot \left( \frac{1}{\alpha +\delta +({\langle k\rangle }-2)\lambda }+\frac{1}{2\alpha +2({\langle k\rangle }-2)\lambda }+\frac{3}{2\delta }\right) + \\ 
    &\left( \frac{{\alpha }^{2}}{(\alpha +({\langle k\rangle }-2)\lambda )(\alpha +\delta +({\langle k\rangle }-2)\lambda)} \right) \cdot 
    \left(\frac{1}{{(\langle k\rangle }-1)\lambda } + \frac{1}{\alpha +\delta +({\langle k\rangle }-2)\lambda }+\frac{1}{2\alpha +2({\langle k\rangle }-2)\lambda }+\frac{3}{2\delta }\right)
\end{split}
\end{equation}
}

and for the $\lambda \ll 1$ and $\lambda \ll \alpha,\delta$ regime:
\begin{equation}
\begin{split}
    T^{\ast} = \frac{1}{{2(\langle k\rangle }-1)\lambda} \cdot \frac{1- \left( \frac{1}{2\langle k\rangle} + \frac{\alpha}{\alpha + \delta}\right)}{\left( \frac{1}{2\langle k\rangle} + \frac{\alpha}{\alpha + \delta}\right)^2} + 
    \left( \frac{1}{2\langle k\rangle} + \frac{\alpha}{\alpha + \delta}\right)\cdot  \left( \frac{1}{\alpha + \delta} + \frac{1}{2\alpha} + \frac{3}{2\delta}\right)
\end{split}
\end{equation}.
\end{widetext}
\FloatBarrier
\bibliography{bibliography}

\end{document}